\documentclass[journal]{IEEEtran}
\onecolumn

\usepackage{url}
\usepackage{hyperref}
\usepackage{graphicx}

\begin{document}

\title{Simulating and Reconstructing Neurodynamics with Epsilon-Automata Applied to Electroencephalography (EEG) Microstate Sequences}

\author{\IEEEauthorblockN{{\sc Chrystopher L. Nehaniv}\IEEEauthorrefmark{1} \ \ \ \ \ \ \ \ \ \ \ \  \and \ \ \ \ \ \ \,\,\,\,\,\,\,\,{\sc Elena Antonova}\IEEEauthorrefmark{2}}\\[2ex]
\IEEEauthorblockA{\IEEEauthorrefmark{1}Royal Society Wolfson Biocomputation Research Laboratory,\\
Centre for Computer Science and Informatics Research,
University of Hertfordshire, U.K.}\\[1ex]
\IEEEauthorblockA{\IEEEauthorrefmark{2}Department of Psychology, Institute of Psychiatry, Psychology and Neuroscience,
King's College London, U.K.}}

\maketitle

\begin{center}
\begin{quote}
\hspace{1.1in}{\it  ``.... these waves are claws,
the boat is caught in them,
you can feel it.''}\\
\hspace{1.13in} -Letters between artist Vincent Van Gogh and his brother Theo, 1888
\end{quote}


\begin{figure}[!ht]
\centering
\includegraphics[width=\columnwidth]{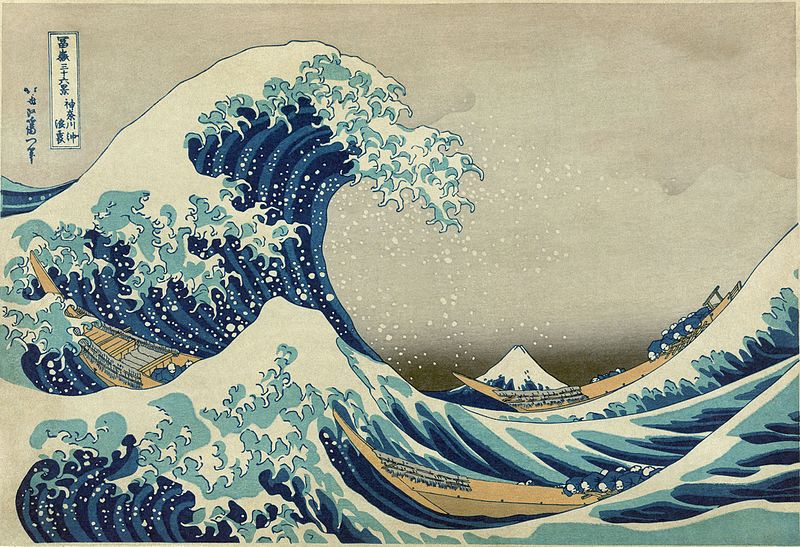}
\caption{{\it Under a wave off Kanagawa} by Hokusai (1760-1849) \cite{Hokusai}. An $\epsilon$-machine 
is a discrete dynamical system which reproduces the temporal behaviour and other characteristics of a process based on rich enough sequences of observations (unlike static
measures and simple statistics).
Thus, like Hokusai's print (also known as {\it The Great Wave}),  an $\epsilon$-machine can be viewed as a portrait capturing in a compact form, the temporal dynamics of
 complex processes, such as the production of sequences of EEG microstates that can be observed in the spatiotemporal neurodynamics of
a particular mental state.}
\label{stillness}
\end{figure}
\end{center}

\newpage

\begin{abstract}
We introduce new techniques to the analysis of neural spatiotemporal dynamics via applying $\epsilon$-machine reconstruction to electroencephalography (EEG) microstate sequences. Microstates are short duration quasi-stable states of the dynamically changing electrical field topographies recorded via an array of electrodes from the human scalp, and cluster into four canonical classes. The sequence of microstates observed under particular conditions can be considered an information source with unknown underlying structure. $\epsilon$-machines are discrete dynamical system automata with state-dependent probabilities on different future observations (in this case the next measured EEG microstate). They artificially reproduce underlying structure in an optimally predictive manner as generative models exhibiting dynamics emulating the behaviour of the source. Here we present experiments using both simulations and empirical data supporting the value of associating these discrete dynamical systems with mental states (e.g. mind-wandering, focused attention, etc.) and with clinical populations. The neurodynamics of mental states and clinical populations can then be further characterized by properties of these dynamical systems, including: i) statistical complexity (determined by the number of states of the corresponding $\epsilon$-automaton); ii) entropy rate; iii) characteristic sequence patterning  (syntax, probabilistic grammars); iv) duration, persistence and stability of dynamical patterns; and v) algebraic measures such as Krohn-Rhodes complexity or holonomy length of the decompositions of these. The potential applications include the characterization of mental states in neurodynamic terms for mental health diagnostics, well-being interventions, human-machine interface, and others on both subject-specific and group/population-level.

\end{abstract}

\IEEEpeerreviewmaketitle

\section{Motivation and Background}

\IEEEPARstart{W}e  focus on using the `grammar' of  electroencephalography (EEG) microstate sequences to distinguish different mental states and their characteristics by employing computational automata- and computer algebraic theoretic tools. 
EEG microstates are quasi-stable maps of electric potential landscapes of short duration, typically classed in a small number of discrete types analogous    to `letters' of an alphabet.  Aberrant tendencies for simple EEG microstate transitions between certain pairs of these discrete types have already been shown to characterize clinical populations (people with schizophrenia or frontotemporal dementia vs.\ healthy controls \cite{Lehmann2005,Nishida2013}). Our research targets finer understanding of temporal neurodynamical structure and patterns using novel computational methods to harness their potential in distinguishing different mental states and in characterizing mental state dynamics in health and disease. 
Specifically, we begin to validate and develop $\epsilon$-machine and algebraic automata methods toward their potential application in clinical diagnosis and mental illness prevention, as well as in promotion of wellbeing.

After reviewing motivation and background in this Section, in Section~\ref{secII} we introduce $\epsilon$-machines (also called $\epsilon$-automata \cite{CrutchfieldYoung1989,Crutchfield1993}), their connection to grammars, and how they can be used to reconstruct unknown underlying dynamics of a discrete dynamical process; in Section~\ref{Exp} we validate the methods in some simple toy models and then on data derived from the study of EEG microstate transitions in schizophrenia patients vs.~controls at a population level; in Section~\ref{Med} we exhibit and compare two  very different $\epsilon$-machines modelling the EEG microstate neurodynamics of an individual meditator and   non-meditator.  Section~\ref{Conc} summarizes our conclusions and discusses possible methodological and real-world applications.

\subsection{``Atoms of Thought'' ?}
William James \cite{James} spoke of `stream of consciousness' -- a continuous flow of thoughts, feelings, and
sensations constituting our experiential `now'. However, modern psychophysiological research employing EEG
has suggested that the transitions between moments of awareness are discontinuous and characterized by
rapid changes in topographic distribution of brain electrical activity followed by periods of quasi-stable spatial distribution referred to as
EEG microstates \cite{Lehmann1998}.  These have been suggested to reflect candidate  `atoms of thought'  that fit together in complex ways in human brain processes and cognition~\cite{Lehmann1998}.

\subsection{EEG Microstates: Topography, Function and Syntax}\label{EEG}

EEG microstates are typological maps of the momentary spatial distributions of electric potential that remain quasi-stable in the sub-second time range and are separated by rapid configuration changes \cite{Lehmann1987}. They can be obtained from a task-related (ERP) or resting-state (EEG) data and further analysed by a number of topographic classes, duration, occurrence, time coverage, and syntax (microstate sequence patterning).  
In terms of topography, most resting-state EEG studies (\cite{Koenig2002,Lehmann2005,Kindler2011})
reveal 4 microstate classes, accounting for approximately 79\%  of the EEG data variance \cite{Koenig2002}.  Two of these classes have asymmetric topography: class A (right-frontal to left-posterior) and class B (left-frontal to right); and two have symmetric topography: class C (frontal to occipital) and class D (frontal to occipital, but with more midline frontal activity than class C). 
Mean microstate duration during resting-state EEG is in the range of 70 to 125 milliseconds (\cite{Lehmann1987,Lehmann1998,Lehmann2005,Koenig2002}).
 The duration, however, and other class parameters vary depending on context and conditions.  Thus, the duration, occurrence, and time coverage of different microstate classes have a developmental trajectory \cite{Koenig2002}.  Alterations of microstate parameters have been reported in schizophrenia 
(\cite{Koenig1999,Lehmann2005,Nishida2013}), depression \cite{Strik1995} and
Alzheimer's disease (\cite{Dierks1997,Nishida2013}). They also vary as a function of wakefulness
\cite{Brodbeck2012}, personality predispositions \cite{Schmidtke2004}, and
hypnosis \cite{Katayama2007}.

In terms of functional significance of the four microstate classes, most recent evidence suggests that their discrete spatiotemporal patterns are signatures of the global neural integration processes associated with conscious cognition and perception.  A simultaneous EEG-fMRI study \cite{Britz2010} found significant correlations between four resting-state EEG microstate classes and four fMRI resting-state networks: class A was correlated with negative blood-oxygen-level dependent (BOLD) contrast imaging signal in areas implicated in phonological processing; class B was correlated with negative BOLD signal in areas implicated in visual processing; class C with positive BOLD signal in saliency and interoceptive networks; and class D with negative BOLD signal in  right-lateralized networks associated with attention reorientation. 
Milz et al. \cite{Milz2016} directly tested the four microstates classes during verbal and visuo-spatial processing and found microstate class A to be associated with visual and class B with verbal processing.  
However, although relative occurrence of classes A and B was significantly different in visual and verbal conditions respectively, all microstate classes occurred during both conditions with the absolute difference in coverage being very small \cite{Milz2016}.  The authors suggest that while increased duration, occurrence, and coverage of EEG microstate classes might be associated with modality-specific processing, they cannot be reduced to these functions. Instead, understanding finely-tuned interplay between the four EEG microstate classes might be necessary for establishing their functionality and their associations with cognition, perception, and subjective experience.  

Studies that explored microstate syntax suggest that a sequence of microstates might indeed be informative.  Basic properties of microstate `language' could include: (1) transition probabilities between pairs of microstates (`syllables'); (2) occurrence probability of short characteristic sequences (`words'); and (3) higher level structure and regularities (`sentences' and `grammar').  Transition probabilities between pairs of microstates (1) have been considered in several studies 
(\cite{Brodbeck2012,Wackermann1993,Lehmann2005,Schlegel2012}).
The importance of short sequences of microstates longer than a single transition (3) is demonstrated by  Lehmann et al \cite{Lehmann2005} showing differences in directional predominance of  microstate transitions between first-episode schizophrenia patients and controls, with predominance of $A \rightarrow D \rightarrow C \rightarrow A$ in patients vs.\ $A  \rightarrow C \rightarrow D \rightarrow A$ in controls.
   Van De Ville et al \cite{VanDeVille2010} presented evidence of higher level structure and regularities in microstate sequences (3) by revealing scale-free dynamics in EEG microstate sequences and concluding that ``modeling microstate syntax needs to go beyond short-range interactions such as modeled by $n$-step Markov chains''. 
Could the formal syntax of such characteristic transitions be detected for more complex sequences?  

This question leads to natural hypotheses that could be investigated using the methods proposed here. Elaborating on the idea that the syntax of the microstate sequences is likely
to be related to the structure of cognition, 
we hypothesize that: (H1) different mental functionalities could entail differential  sequences of brain processes resulting in different characteristic microstate sequences or syntax.
 As a simple case, this includes different transition probabilities between microstates in different context, e.g. in a visualization task if attention momentarily wanders, one would predict microstate D associated with attention-reorienting would be more likely to
be followed by one associated with visual activity than verbal.  
(H2) Longer microstate sequences (`words') with constrained structure would be expected to reflect more coverage by contextually relevant microstates, in particular those associated with the task at hand.
(H3) Extracting short microstate sequences characterizing different brain processing steps during various tasks or mental processes could reveal whether there are (i) generic steps that occur in cognition as well as (ii) task- or context-specific ones for the given mental process. Since all 4 EEG microstate classes derived by Koenig et al.~\cite{Koenig2002} occur in various spatial- and object- visualization and verbal tasks  \cite{Milz2016}, 
we would predict that generic cognitive steps would be associated with common microstate syntax (e.g.~framing by the same short sequences), and differences in task and context would be characterized by the differences in microstate coverage and syntax.

Syntactic structure is well studied in formal language theory where different classes of automata models act as generators or recognizers of a `language' of possible sequences (e.g. \cite{Hopcroft}, \cite{EilenbergB}, 
\cite{Nehaniv1996}, \cite{wildbook}).  To study microstate syntax, such generators/recognizers can be constructed from microstate sequence data using $\epsilon$-machines methodology, developed for inferring the causal structure of processes from a sequence of discrete measurements or observations.  Unlike fitting parameters to the architecture of a pre-given model (e.g. \cite{Gaertner2010}), the structure of an $\epsilon$-machine is derived from the data. The $\epsilon$-machine construction infers a minimum number of causal states necessary to optimally predict future observations \cite{CrutchfieldYoung1989,Shalizi2003}. The causal states of $\epsilon$-machines collect trajectories of the process up to the present moment that probabilistically predict exactly the same possible futures, and, with sufficient data, asymptotically give complete information of the syntax of the process. These discrete dynamical systems allow not only to address such syntactic characteristics of neurodynamics, but to determine meaningful underlying computational architectures of processes that could produce the observed sequences and to characterize their complexity  \cite{Crutchfield1993}.

\section{$\epsilon$-machines, Grammars, and Algebraic Invariants}\label{secII}

Unknown dynamical processes produce sequences of discrete observations from a finite set of possibilities (`alphabet') corresponding to `letters'  (e.g. the four canonical EEG microstates $\{A,B,C,D\}$  characterized by
Koenig et al.\ \cite{Koenig2002}), and to attempt to capture these processes
$\epsilon$-machines are generative dynamical models determining classes of formal languages over the alphabet and a probability measure on observed words (sequences or time series of observations).   It will be assumed the process being studied is {\em stationary}, i.e. governed by the same dynamics at each moment of time,  or at least {\em conditionally stationary}, i.e. the  present dynamics depend at most only on previous history.  Then 
$\epsilon$-machines are provably optimal non-linear recursive predictors for discrete time series  (\cite{CrutchfieldYoung1989,Crutchfield1993,Shalizi2004}).

\subsection{$\epsilon$-Machines}
 
Given discrete time series data, $\epsilon$-machines are minimal {\em sufficient statistics}, i.e. given the past history they allow complete determination of the probability distribution over possible futures \cite{CrutchfieldYoung1989}.  Each past history determines a unique {\em causal state} of the $\epsilon$-machine and two past histories determine the same causal state if and only if they determine the same probability distribution over possible futures. Moreover, causal states have a deterministic automata structure:
given a past history $h$ and next observation $x$ of the time series, the new causal state is the causal state of history $hx$, that is, the history where $h$ then $x$ occurred.
The causal states are minimal in that any other statistic that determines that determines possible futures given past history must contain at least as much information as the causal states  \cite{CrutchfieldYoung1989} and so can be mapped onto the them.

An $\epsilon$-machine (or $\epsilon$-automaton) is a formally specified automaton ${\cal A}=(Q,X,\delta, P)$,  often constructed from observation sequences
\cite{Crutchfield1993}.  It has a finite alphabet $X$ from which {\em observations} are drawn, a set  $Q$ of underlying {\em causal states}, and a deterministic
transition map $\delta: Q \times X \rightarrow Q$, giving the next state $q'=\delta(q,x)$ if the automata was in state $q$ and observation $x\in X$ was
observed.  To each state $q\in Q$ is assigned a probability distribution $P_q$, giving the probability $P_q(x)$ of the next observation
being $x$ when the machine is in state $q$.   
$$ \sum_{x\in X} P_q(x) =1, \mbox{ for all $q \in Q$}. $$

A {\em trajectory} of $\cal A$ starting in state $q=q_0$ is a sequence of states $q_0, \ldots q_{\ell}$ and observations $x_1,\ldots, x_{\ell}$ with
$q_{i+1}=\delta(q_i,x_i)$ for all $0\leq i < \ell$, for some {\em length} $\ell \geq 0$.  The probability of $w = x_1 \ldots x_{\ell}$ occurring in state $q=q_0$ 
is  thus $$P_q(w)=\prod_{i=0}^{\ell-1} P_{q_i}(x_{i+1}).$$    Of course, also
$\sum_{w \in X^{\ell}} P_q(w) =1$.

The {\em behaviour} of the $\epsilon$-machine is to iteratively emit a letter $x$ with probability $P_q(x)$ when in state $q$ and then to transition to new state $q'=\delta(q,x)$.
Here time is modelled as discrete and defined by events (or observations), i.e. the index of time $t$ is regarded as integer-valued and the next moment of time $t+1$
is determined whenever $x_{t+1}$ is observed.\footnote{Note that if $P_q(x)$ is zero, it is sensible to regard $\delta(q,x)$ as undefined, as $x$ can never be observed in
any transition from state $q$. Alternatively, 
one could set  $\delta(q,x)= *$,  an adjoined `impossible' state in this case, i.e., $*$ is a `sink state', which can 
never be left once entered. That is,  $\delta(*,x)=*$ for all $x\in X$. }

At each state $q$ and sequence length $\ell$ an $\epsilon$-machine determines a possible set of observation sequences
 $$L_q^{(\ell)}=\{ w \in X^{\ell} : P_{q}(w) \neq 0 \},$$  as those occurring from state $q$ with non-zero probability.   The {\em future morph} at
 state $q$ is the union of all possible observation sequences  possible at $q$, i.e.
  the set $L_q=\bigcup_{\ell > 1} L_q^{(\ell)}$ where each $w \in L_q$ has an associated probability $P_q(w)$. If $|w|=\ell$,
  we have probability $P_q(w)$ over all sequences of length $\ell$ of observing $w=x_1,\ldots, x_{\ell}$ from causal state $q$. 
  The future morph at state $q$ is  thus the set of all possible futures with particular probabilities associated to them. 

Suppose the $\epsilon$-machine has so far generated a (finite or infinite) sequence 
$\ldots x_{t-m} \ldots  x_{t-1} x_t$  up to the present moment (indexed by $t$) and is in state $q$.  
Observe that in an $\epsilon$-automata for each state $q$, the future morph is conditionally independent of the past.
That is,  the probability of any sequence  $w$ of observations of length $\ell$  occurring at $q$ 
depends only on the current state $q$ and not on any details of the sequence $\ldots x_{t-m} \ldots  x_{t-1} x_t$.

\subsection{Physical Clock Time vs. Event-Driven Time }

For time as modelled in $\epsilon$-machines, making a new observation defines a `clock-tick' and determines a transition to the next causal state. 
We may choose to record an observation  according to a regular interval of
physical clock time, e.g. recording the sequence of observations $x_1 x_2 x_3 \ldots ... $  at 250 Hz  with 4 milliseconds (ms) between observations, in our case EEG microstates.   
Alternatively, we may choose a different sampling rate, e.g. 100 Hz, so that 10 ms elapse between observations.
We may also choose `event-driven' time instead, with the $\epsilon$-machine making a transition whenever  an observation is made, and only then.
 Moreover, we could allow a mixture of physical time and event-driven time: e.g. the $\epsilon$-machines receives an observation only if the observational reading has changed (or a maximal time interval
 has elapsed);  in this case $x_i \neq x_{i+1}$ (unless the maximum time has elapsed),
 and the shortest time interval interval between
 the two observations $x_i$ and $x_{i+1}$ is the sampling rate.\footnote{   For example, if the sampling rate is 500 Hz (one
 reading per 2 ms) and the maximal time is set to 50 ms, then observation $x_{i+1}$ could occur at some multiple
 of 2 ms after observation $x_i$, at the first reading where the observations differ but not more than 50 ms later.}
In every case, time  in $\epsilon$-machines is always indexed by the arrival of discrete observations. 
 
 \subsection{$\epsilon$-machine Reconstruction from Observations}\label{Epsilon}
 
 Crutchfield and colleagues (\cite{Crutchfield1993,CrutchfieldYoung1989,Shalizi2003,Shalizi2004}) describe methods to approximately reconstruct an $\epsilon$-machine
 from a given sequence or sequences of observations of a physical process that is conditionally stationary.
 
A partition of a set of all finite sequences $w$ of observations  is said to make the future conditionally independent from the past if
 for all $m>0$ the probability of each $w' \in X^{m}$ occurring after each member of the partition class $[w]$ is the same. Members of a partition class $[w]$
 are regarded as {\em equivalent histories}. That is, for any two observation sequences $w$ and $w'$ in $[w]$,
 the probability of any future sequence of observations $x_1, \ldots, x_m$ following $w$ is the same as the probability of the future sequence of
observations $x_1, \ldots, x_m$ following $w'$.

Given such a partition, we can define an $\epsilon$-machine with alphabet $X$ (the set of possible discrete observations), states given by 
the partition classes $[w]$, and  transition function $\delta([w],x)=[wx]$, for each $x \in X$ .   
The transition function is well-defined by the condition that the partition makes the future conditionally independent of
the past. It follows that the probability $P_{[w]}(x)$ that $x$ is observed after the having just observed $w$ is independent of
the representative $w$ of the partition class $[w]$, so is well-defined to. Since the transition function is defined in terms of states and observations,
it turns out that  it suffices to consider just the next observation after word $w$ in $\epsilon$-machine reconstruction  \cite{Shalizi2003}.
 
 Due to the limitations of the finiteness in real data sequences of discrete observations, a  maximal window size $k$ is selected as a parameter of
 the method, and all strings of actual observations of length at most $k$ are considered in estimating such a partition on words of length at most $k$.   
  Based on observational data, one approximates the causal state set by starting with all   
sequences of observations of length zero, and the partition of all sequences into a single universal class partition class,  
 where observation $x\in X$ occurs with probability its observed frequency in the data.
 Supposing that the partition has been constructed up to a certain stage with
 all words in partition classes having length at most $m$ ($m < k$),  the one-step probability distribution over observations
 for each partition class $[v]$
is estimated by simple maximum likelihood, to reflect the frequency of observations $x$ appearing after any histories in $[v]$ in the data.
 Extending each $w$ of length $m$ in a partition class $[w]$ by a letter $y$ (one-step of previous history) to get $yw$, one checks whether in the one-step probability distribution over observations following $yw$ is the same as that of partition class $[w]$, or, if not, of any other $[v]$ obtained so far.  (Statistical tests are applied to order to decide whether the distribution over next observations of $yw$ and $[v]$ are the `same' in software implementations \cite{Shalizi2004}.)  If so, then $yw$ is added to $[v]$ (which may well be $[w]$).  If not, a new partition class $[yw]$ containing $yw$  is created. In practice $yw$ needs only be considered if it occurs in the observation sequence(s) from data. 
 One continues this for all letters $x \in X$ and all partition classes found so far,
 using only words up to length $k$ as derived from data.   The transition function is defined by $\delta([w],x)=[wx]$ if $|wx|\leq k$ or by the $[(wx)_k]$ otherwise, where $(wx)_k$ denotes the last $k$ letters of $wx$.   One checks whether $\delta$ is well-defined for $w$ and $x$, i.e. independent of the representative $w$ of $[w]$. If not, one iteratively partitions  $[w]$ into new states to make it well-defined.  This causal state splitting process is iterated for all candidate causal states $[w]$, and is guaranteed to converge due to finiteness of the data, with a worse case run-time of $O(|X|^{2k+1})+O(N)$ where $N$ is the number of observations in the data \cite{Shalizi2003}.
 Examples illustrate success of this reconstruction even when the process modelled is non-Markovian \cite{Shalizi2003}.

   \subsection{Deriving Grammars from $\epsilon$-machines}
If the $\epsilon$-machine is finite, we obtain a
(probabilistic) regular grammar for a formal language $L_{q_0} \subseteq X^*$ for each causal state $q_0$ of the $\epsilon$-machine. 
  (See e.g. 
 \cite{Hopcroft}
 for an introduction to formal languages.) 
We have a variable symbol $V_q$ for each causal state $q \in Q$, a terminal symbol $x$ for each observation letter $x\in X$. 
For each causal state $q$ of the automata and observation letter $x\in X$, we have a  grammar rule 
$$V_q \rightarrow x V_{\delta(q,x)} $$
 if $\delta(q,x)$ is defined, i.e.,  $P_q(x)>0.$
We also add rules 
$V_q \rightarrow  \lambda$, the finite string with no letters; that is, $V_q$ may be replaced by the empty string terminating the generation of a finite word.
 Then $L_q$ is the language of all strings in the terminal symbols that can be reached using these rules starting from symbol $V_q$.  Observe that we may assign
 probabilities $P_q(x)$ to the first kind of rule above, in which case the probability of $w \in X^{\ell}$ being generated is $P_q(w)$ as defined above.

  Note: this derivation of a probabilistic grammar will work even if the number of 
 causal states is infinite, although the languages $L_q$ need not then be rational (i.e. recognizable by a finite-state automaton). 
 \subsection{Invariants of Mental States and Complexity Measures}

As reviewed in Section \ref{EEG}, the parameters of  EEG microstates and their sequences such the duration or coverage of given EEG microstate classes, or the pairwise transition probabilities
 from one microstate class to another have been shown to distinguish between 
 between different mental states (e.g. visual, verbal, or interoceptive information processing, and  attention reorientation) \cite{Milz2016}, as well as clinical populations (e.g. schizophrenia, frontotemporal dementia, and Alzheimer's disease) and healthy controls (\cite{Lehmann2005,Nishida2013}).
 
 The $\epsilon$-machine construction allows much richer, predictive  {\em measures of a system's behaviour by specifying its dynamics in a more complex way} than has been applied previously in EEG studies. To each observational epoch or a set of epochs, the construction
  associates (1) a deterministic finite automaton
 with probabilistic transitions for each letter from each state, (2) a probabilistic regular grammar, and (3) probabilities to particular sequences of
 observations. The $\epsilon$-machine can be used to derive a number of other measures, including statistical complexity (the $\log_2$ number
 of states of the automata), entropy rate,  Krohn-Rhodes complexity  (minimal number of permutation computing levels in a cascade decomposition) of these automata and their algebraic invariants such as
 their simple subgroup divisors (`atoms' of computation), their natural subsystems  (\cite{wildbook,royalsoc}), and the length of their holonomy decomposition (\cite{EilenbergB,AttilaJamesChrystopher}).
 We compute these using computer algebraic software SgpDec \cite{AttilaJamesChrystopher}.
 These measures can be used as dependent variables in experiments to distinguish different mental states or different clinical populations,
 and could potentially be applied in diagnostic tools.

\section{Simulation Experiments}\label{Exp}

\subsection{Simulating the EEG Microstate Sequences Corresponding to a Mental State} 

For the purpose of the simulation experiments presented in this section 
and in the $\epsilon$-machine analysis of actual EEG data of Section IV, we use sequences of an (arbitrary) length of 30,000 EEG microstates. This is the number of EEG microstates derived, for example, from a human participant during  a particular experimental condition that probes/engages a certain  mental state or process if one uses two-minute interval  epochs during which neurodynamics are assumed to be conditionally stationary, at a recording rate of 250 Hz per microstate (which is a standard rate used in the field).
Using 
$\epsilon$-machine reconstruction as outlined in Section~\ref{Epsilon}, we can generate a discrete dynamical system model that is optimized to produce sequences of microstates statistically indistinguishable from the observed microstate sequence data and study its various characteristic properties. 
 (Section~\ref{Med} gives $\epsilon$-machine
reconstructions for two individuals, an experienced meditator and a non-meditator, from actual EEG microstate sequences of exactly this kind.)

To validate our methods we  first test the capacity of $\epsilon$-machines to reconstruct known underlying processes using simulated data.
 We describe
the grammatical structure of processes using probabilistic regular grammars,  test the capacity for $\epsilon$-machine reconstruction to
recover the generating automata, and apply computer algebraic automata analysis using a computer algebra GAP extension by the first author
 to CSSR \cite{Shalizi2004} and SgpDec \cite{AttilaJamesChrystopher}.

For the simulations, we use as a departure point the parameters such as observed frequencies of occurrence and pairwise transition probabilities for different EEG microstate classes,  with parameters derived from clinical experiments used in \ref{ExpD}.
In the rest of this Section, we carry this out for synthetic observational data, (1) with no preferred sequence structure, (2) with deterministic and non-deterministic cycles, and (3) with structure based on observed transition probabilities between microstates from a clinical study.

\subsection{Experiment 1 : No Preferred Sequences}

Suppose that observations show any sequence of length $n$ as equally likely.  It follows that given any sequence of observations $w$,
all words $v$ of length $\ell$ occur with equal likelihood following $w$. Each observation $x \in X$ following $w$ is equally likely, 
and the probability of $wv$  is the same as the probability of $wv'$ if $|v|=|v'|$.  Thus the $\epsilon$-automaton has exactly one causal state $q_0=[w]$,
and the probability of $x$ in state $q_0$ is $\frac{1}{|X|}$, and the transition function is the trivial one $\delta(q_0,x)=q_0$ for all $x \in X$. Its entropy rate is $\log_2 |X|$ and
statistical complexity is zero. The corresponding grammar has one variable symbol $V_0$ corresponding to the unique sate of the $\epsilon$-machine and 4 equiprobable rules:
$$V_0\rightarrow AV_0,\ V\rightarrow BV_0, V_0\rightarrow CV_0, V_0\rightarrow DV_0.$$
Similarly, if the microstates $A,B,C,D$ occur with probabilities $p_A, p_B, p_C, p_D$ and these are independent of all previous history, then
the four rules would have these respective probabilities.  Straightforward computational experiments confirm $\epsilon$-machines easily recover this structure.

\subsection{Experiments 2a and 2b: Cycles}
\begin{figure}[!h]
\begin{center}
 \includegraphics[scale=.33]{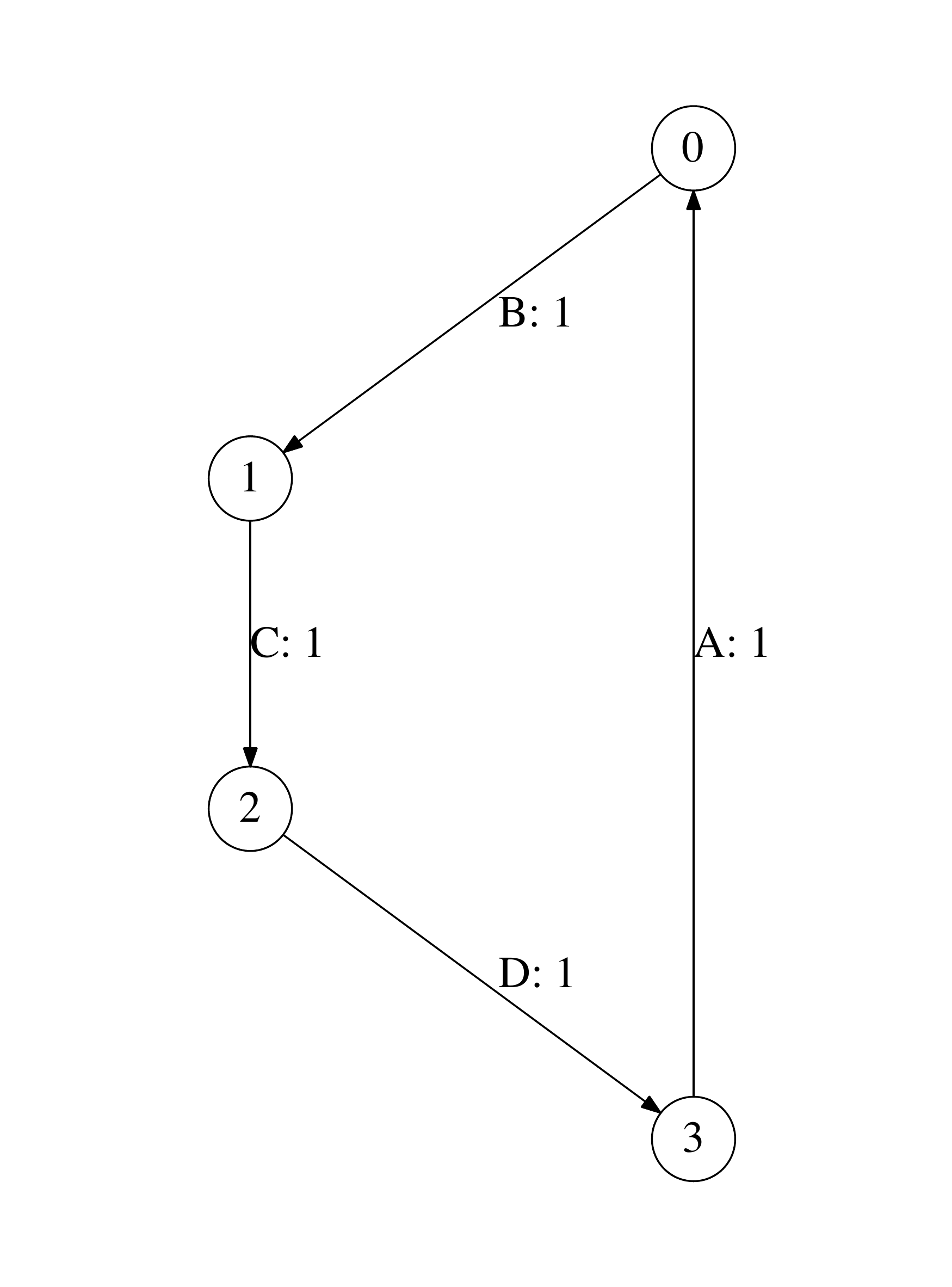}
  \includegraphics[width=.44\columnwidth]{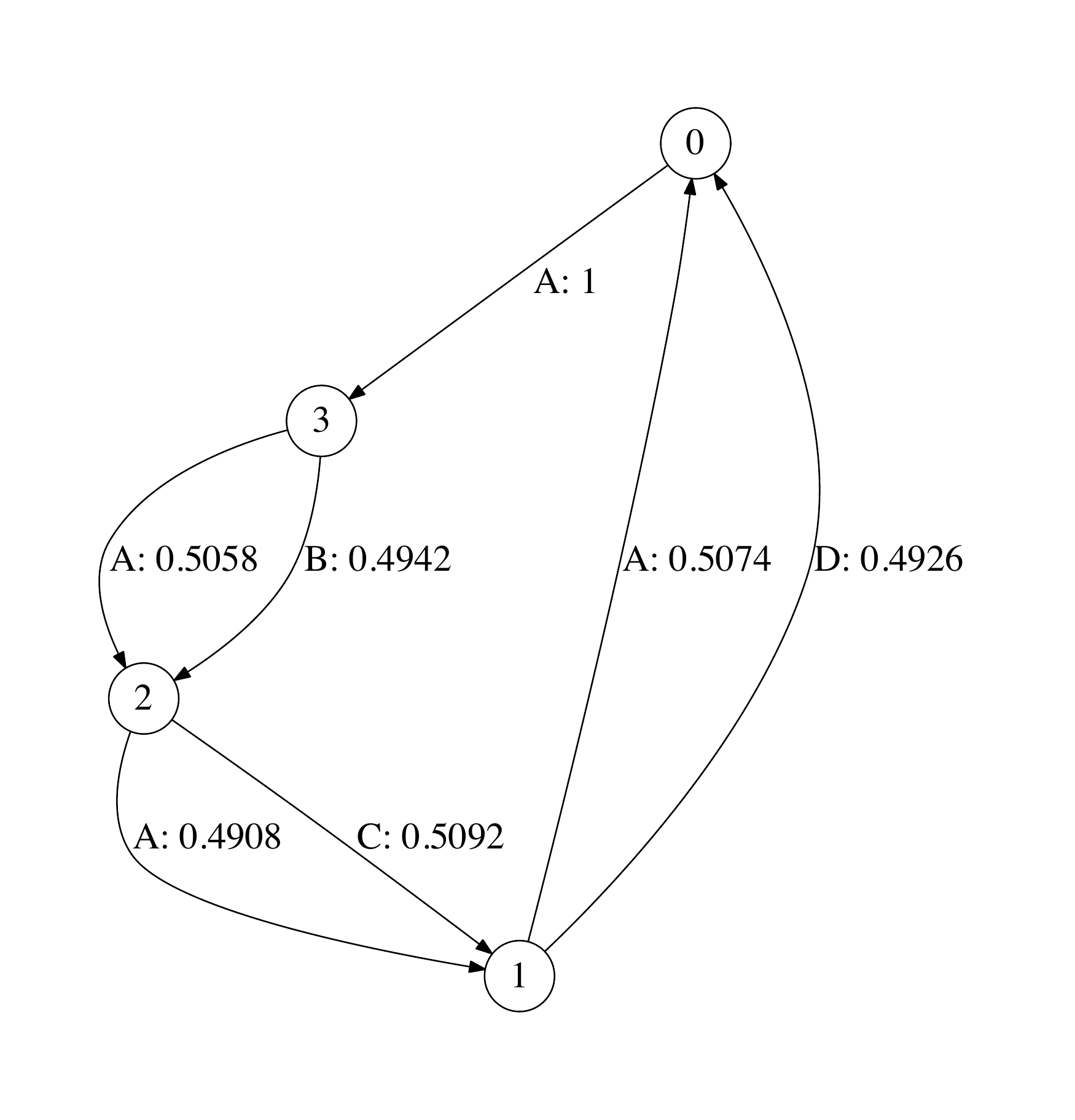}
 \caption{Reconstructed $\epsilon$-machines from simulated data in Experiments 2a (left) and 2b (right). Causal states have been reconstructed perfectly. Arrows
 label transitions with the next observed EEG microstate and transition probability. 
These probabilities have been recovered from the synthetic data sequences perfectly in Exp.~2a and to within $\sim 0.7$\%  in Exp.~2b.
Statistical complexity is 2, and entropy rates are $0$ (left) and ~$0.75$ (right). The Exp.~2a yields an $\epsilon$-machine with a transformation semigroup of size 17 generated by the 4 partial constant mappings corresponding to the observed microstates. The holonomy decomposition has two levels and the Krohn-Rhodes complexity is zero, as all subgroups are trivial. 
 The $\epsilon$-machine for Exp.~2b is interesting since the semigroup transition corresponding to $A$ permutes the causal states generating a cyclic group of order 4, so Krohn-Rhodes complexity is non-trivial, and the holonomy decomposition has two levels (see \cite{wildbook} or \cite{AttilaJamesChrystopher}) with the
 computing group level on top. The semigroup has 21 elements and is generated by the order 4 permutation of the states corresponding to observing $A$ and three partial constant maps corresponding to transitions $B$, $C$, $D$. 
 \label{cycles}}
 \end{center}
 \end{figure}

Given a cycle of observations, $A, B, C, D$, which repeats indefinitely (here 90 times), the reconstructed $\epsilon$-machine has 4 states and probability 1 to emit the next letter in the sequences (Figure~\ref{cycles}, left). 

A  variant is to repeat the cycle $A, B, C, D$ indefinitely, except that, with probability $p=\frac{1}{2}$, $A$  is observed rather than $B$, $C$, or $D$ in any position where these letters would otherwise occur in the repeating sequence.   An artificial sequence of length 30,000 simulated EEG microstates was generated following this rule with the $\epsilon$-machine
successfully reconstructing the structure of the underlying process (Figure~\ref{cycles}, right).

 These two $\epsilon$-machines have a particularly simple structure, with 4 states  $Q=\{q_A,q_B,q_C,q_D\}$, which record the
the most recently observed EEG microstate, and transition function $\delta(q_X, Y)= q_Y$ for all $X, Y \in \{A,B,C,D\}$.
Any EEG microstate sequence could be crudely modelled in this way,   where  the probabilities $P_{q_X}(Y)$  are derived from
empirical measurement.  (Both $\epsilon$-automata recovered
in Experiment 2a and 2b are of this form.) Such a model exhibits no dependence on the history of the dynamics beyond the most recent microstate,
which might {\em not} reflect underlying neurodynamics sufficiently well.

\subsection{$\epsilon$-machines of EEG microstate sequences in Schizophrenia Patients vs.\ Healthy Controls}\label{ExpD}

Lehmann et al. \cite{Lehmann2005} reported the percentages of transitions between distinct EEG microstates for schizophrenia patients vs.\ healthy controls, and established that   
significant differences exist between the two populations in directional predominances of certain transitions between microstates, and also
 a predominance of EEG microstate sequence
 $ ADCA$ over $ACDA$ in patients and the reverse in controls.
 
Though the Lehmann et al. (2005) publication does not contain full microstate sequences, the percentages of occurrence of each type of transition between distinct microstates reported for the two groups  of 27 patients and 27 controls (Table 3(A), p.~150 \cite{Lehmann2005})  allowed us to
calculate the probability distributions at {\em group level} of each distinct microstate following a given microstate $A$, $B$, $C$ or $D$.  
Using the reported percentages for each transition between microstate pairs of type $x\rightarrow y$ ($x\neq y$), we calculated that probabilities of
the next {\em distinct} microstate in schizophrenia {\em patients} given the current EEG microstate were as follows:\\

\begin{center}
{\small
$\begin{array}{lcccc}
\mbox{probability}& \mbox{to $A$}& \mbox{to $B$}&  \mbox{to $C$} &  \mbox{to $D$}\\
\mbox{from $A$:}&  0 & 0.275319 & 0.391489 & 0.333191 \\
\mbox{from $B$:}& 0.337513 & 0 & 0.333501 & 0.328987 \\
\mbox{from $C$:}& 0.322104 & 0.225507 & 0 & 0.452389\\
\mbox{from $D$:}& 0.270644 & 0.248818 & 0.480538 & 0
\end{array}$\\
}
\end{center}
and  in {\em controls} were as follows:
\begin{center}
{\small
$\begin{array}{lcccc}
\mbox{probability}& \mbox{to $A$}& \mbox{to $B$}&  \mbox{to $C$} &  \mbox{to $D$}\\
\mbox{from $A$:}& 0 & 0.296104 & 0.390649 & 0.313247 \\
\mbox{from $B$:}& 0.244558 & 0 & 0.329065 & 0.426376 \\
\mbox{from $C$:}& 0.243999 & 0.281978 & 0 & 0.474024\\
\mbox{from $D$:}& 0.228202 & 0.335831 & 0.435967 & 0
\end{array}$\\
}
\end{center}

(Note: rows -- but not columns -- sum to $1$ in these tables.)

 We attempted to simulate corresponding EEG microstate sequences for patients and controls using simple $\epsilon$-machines with
no dependence on history except for the current microstate.
The tables allow the construction of simulated microstate sequences with the probabilities of the next observation depending only the current microstate.
In effect, these tables are generative $\epsilon$-machine models with a causal state for the current microstate reading, i.e. with 4 causal states with differing probabilities
over which distinct microstate will occur next (according to the rows above). 
We generated such sequences of length 30,000 microstate transitions according to the above transition probability tables for patients and controls.

\begin{figure}[h]
\begin{center}
Patients \ \ \  \ \ \ \    \ \ \ \ \   \ \ \ \ \  \ \ \ \ \ \ \ \ \ \ \ \ \ \ \ \ \ \ \ \ \ \ \ \ \ \ \ \ \ \ \  \ \ \ \ \ \ \ \ \ \ \ \ \ \ \ \ \ \ \ Controls\\
 \includegraphics[width=.49\columnwidth]{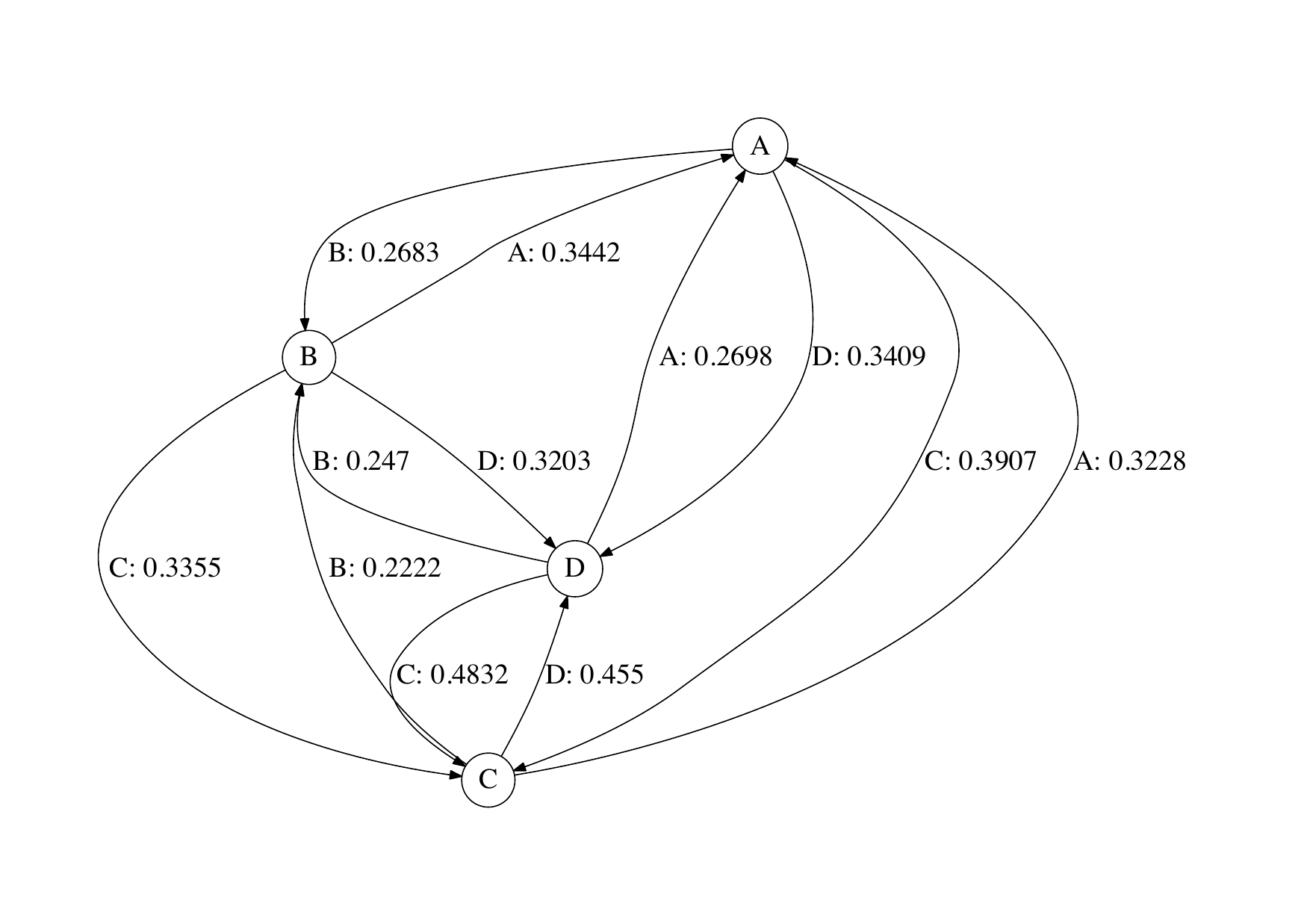}
  \includegraphics[width=.49\columnwidth]{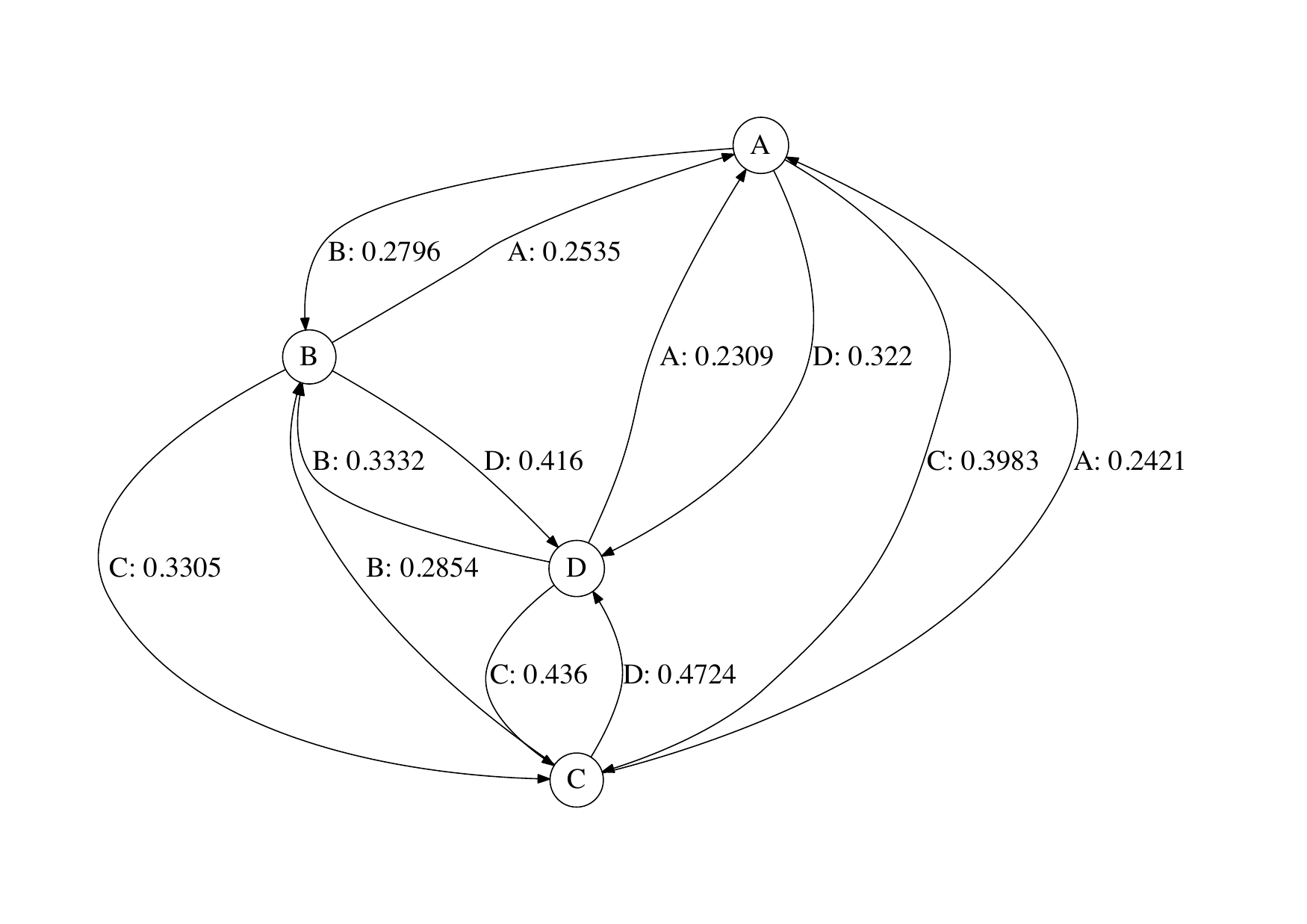}
   \end{center}
 \caption{Reconstructed $\epsilon$-machines from simulated EEG microstate sequences for Schizophrenia Patients and Healthy Controls. Casual states in this
 case correspond to the most recent microstate observed. Arrows are labelled by the next observed microstate and the probability of that microstate being observed from the given casual state. Inferred probabilities have been recovered to within $\sim$1\%  of those used in generating the simulated length 30,000 EEG microstate
  sequences. Statistical complexity of both is 2; entropy rates are 1.54438 (patients) and 1.54428 (controls). 
  The algebraic invariants analysis yields that the transformation semigroups of the control and patient $\epsilon$-machines are isomorphic since they differ only in non-zero transition probabilities. Both have 17 elements  (generated by the four partial constant maps to each EEG microstate). They have holonomy decomposition with two identity-reset levels  and contain no non-trivial symmetry groups, hence their Krohn-Rhodes complexity is zero.}
 \label{schizophrenia}
 \end{figure}

Using these simulated microstate sequences,
 we reconstructed $\epsilon$-automata for the patient and control groups as in Figure~\ref{schizophrenia}.   
These inferred $\epsilon$-machines correctly recover the number of states and closely approximate the transition probabilities in the simple model. (Longer simulated 
 microstates sequences would yield further precision in recovering the transition probabilities, already within $\sim$1\%  of the generative models.)
 This further validates the $\epsilon$-machine reconstruction method in that 
the resulting associated probabilistic regular grammars reflect the reported data well, at least at the level of pairwise transitions between microstates.

One might suspect that a four-state model (equivalent to one-step hidden Markov model) is enough to replicate much of the data Lehmann and colleagues report.
A simple transitional structure like this might possibly reflect the neurodynamics of patients and controls (at a group level):
 Indeed, we can analytically 
derive that, if there were only 4 states with the transition probabilities we derived from Lehmann et al.'s work~\cite{Lehmann2005}, then the 
probabilities of the length 4 sequences $ACDA$ and $ADCA$ 
are  1.15\%  and 1.24\%   for patients, and 0.811\%  and 0.640\%  for controls, respectively.\footnote{For the sequence $ACDA$,  one
 calculates its occurrence among microstate sequences of length $4$ by starting at the microstate $A$ with probability equal to its frequency of occurrence times the probabilities of transitioning from  $A$ to $C$, then $C$ to $D$, then $D$ to $A$ for patients and for controls, respectively; similarly for the sequence $ADCA$.
In \cite{Lehmann2005}, microstate $A$ occurs  2.91 times/sec  with 12.12 microstates / sec accounting for 24.01\%  of microstate occurrences  for patients,  
and  microstate A occurs 2.17 times/sec with 11.30 microstates / sec accounting for 19.20\%  of microstate occurrences for controls.
Using the transition probabilities for patients, $24.01\%  \times p_{AC}\times  p_{CD}\times p_{DA}=1.15\% $ of 4-microstate words are expected to be $ACDA$, similarly
$1.24\% $ are expected to be  $ADCA$ for patients,  and $0.811\% $ for $ACDA$  and $0.640\% $ for $ADCA$ for controls. 

Thus in a 30,000 microstate sequence one expects:
342 $ACDA$s and 371 $ADCA$s for patients and 
243 $ACDA$s and 192 $ADCA$s for controls, respectively.
In  our simulated sequences generated considering only the transition probabilities, as described above, 
 for patients $ACDA$ occurred 329 times, while $ADCA$ occurred 389 times, whereas in the model for controls $ACDA$ occurs 235 times and $ADCA$ occurs 194 times,
 close to what is expected.}  
 The analytical estimates are in fairly good agreement with the simulated microstate sequence and in line
with the predominance of $ADCA$ over $ACDA$ in patients and the reverse in controls reported in \cite{Lehmann2005}.

However, starting the cycle $ACDA$ at  $C$ or $D$ instead of $A$ yields $CDAC$ and $DACD$ and Lehmann et al report  these 3 variant cycles ($ACDA$, $CDAC$, and $DACD$) occurring 
 as 2.76\%  of length 4 words in patients, and similarly the 3 cyclic variants of  $ADCA$  (namely $ADCA$, $CADC$,  and $DCAD$)  are reported to
                 occur as 3.47\%  of length 4 words in patients. 
In controls,  the respective percentages     were reported as $2.50\% $ and $2.10\% $.
Using the occurrence frequencies of $A$, $C$ and $D$ from the study and the above transition probabilities, 
         one would expect 3.82\%  for $ACDA$ variants and 4.11\%  for $ADCA$ variants in patients, 
 and  3.22\%  for $ACDA$ variants and 2.54\%  for $ADCA$ variants in controls. 
 While these percentages are somewhat higher than those reported in Lehmann study, both  the original study and the $\epsilon$-machine reconstruction models 
are in agreement on the predominance of $ACDA$ variants in controls and $ADCA$ variants in patients.

The discrepancy between  these analytically estimated simple transition model percentages with the actual percentages reported in the study suggests that
 the EEG neurodynamics  yields a syntax requiring a richer model to capture its dynamics.  A full $\epsilon$-machine
reconstruction from actual data, at both the individual and group levels, could be considerably more complex, and might reveal additional syntactic structure.
 This is a strong argument for using methods sensitive to
higher level grammatical structure such as $\epsilon$-machines to explore EEG microstate sequences in greater depth.

 It would therefore be useful to apply our
techniques to the actual EEG microstate time series and compare the resulting inferred $\epsilon$-machines with the above in order to detect further
possible structural differences between the spatiotemporal neurodynamics of schizophrenia patients vs.\  controls in terms of
process structure, grammar, and algebraic invariants.

\section{EEG Microstate Dynamics of Mind Modes Reconstructed with $\epsilon$-Machines}\label{Med}

In this section, we report $\epsilon$-machine reconstruction of the resting EEG microstate data from two individuals, an experienced meditator and a meditation-na\"{\i}ve healthy control collected  during our pilot project (an unpublished MSc thesis, King's College London) exploring $\epsilon$-machine potential for differentiating the EEG microstate data during three information processing modes or mind modes:  mind-wandering, focused attention, and open presence \cite{GwynThesis}.

\begin{figure}[!ht]
\begin{center}
Meditator: \ \ \ \ \ \ \ \ \ \ \ \ \ \ \ \ \ \ \ \ \ \ \ \ \ \ \ \ \ \ \ \ \ \ \ \ \ \ \ \ \ \ \   \ \ \ \ \ \ \ \ \ \   Non-meditator:\\
 \includegraphics[width=.49\columnwidth]{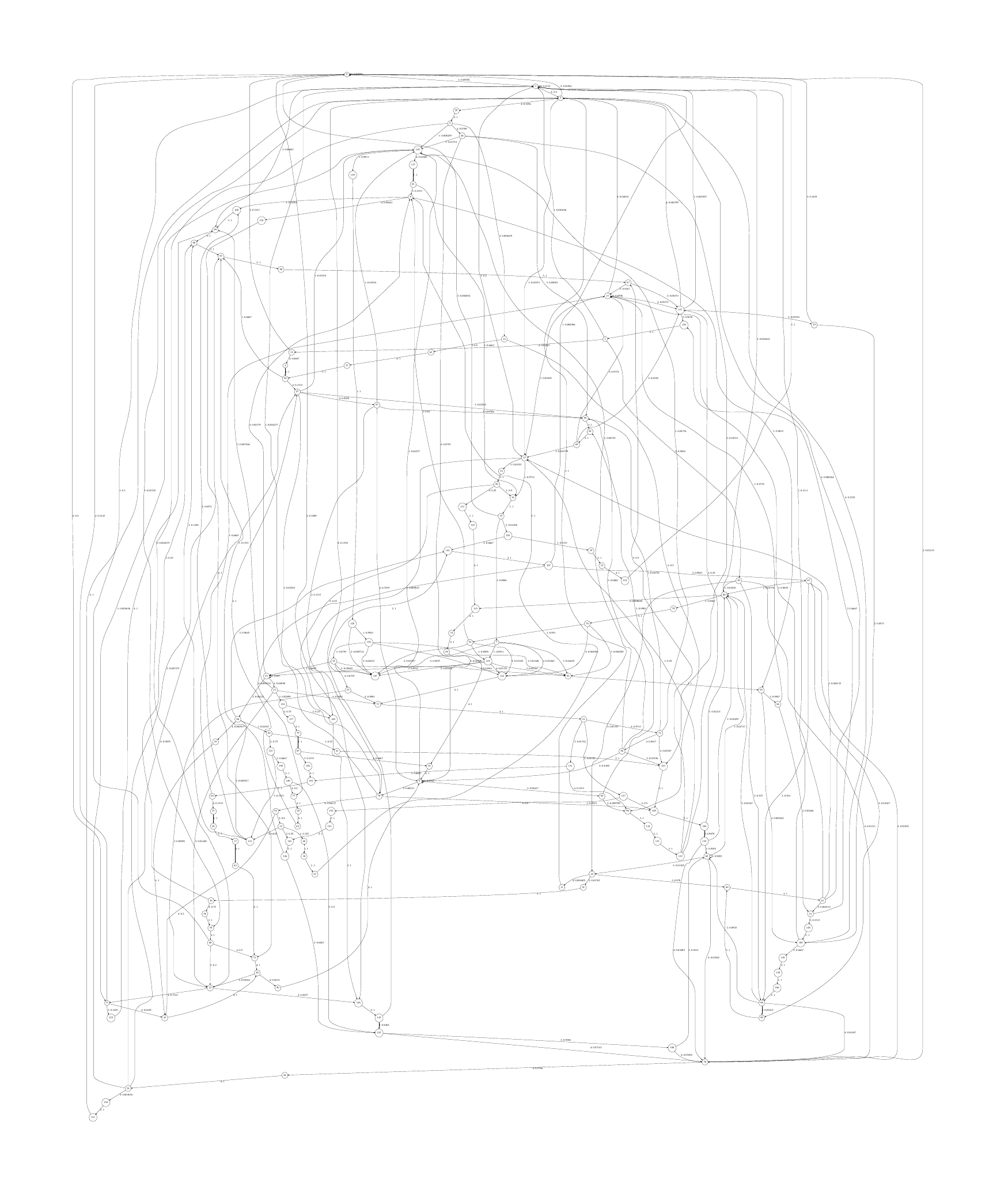}
  \includegraphics[width=.5\columnwidth]{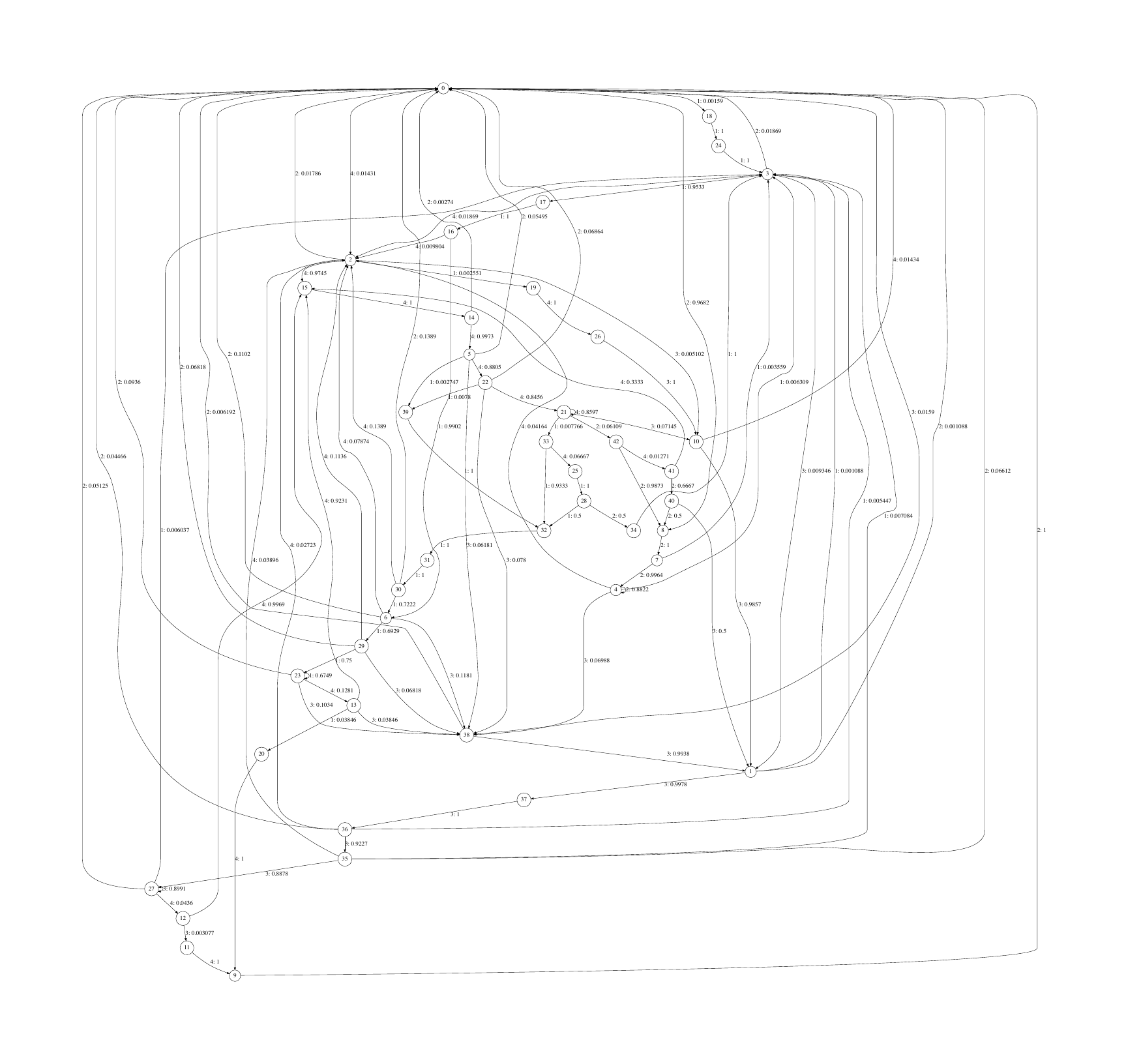}
  \end{center}
 \caption{Reconstructed $\epsilon$-machines from EEG microstate sequences for Meditator vs.\ Non-meditator during mind-wandering.
  The $\epsilon$-machine for the meditator has 153 causal states, transition semigroup with holonomy decomposition length 24 and Krohn-Rhodes complexity $1$, indicating 
presence of a non-trivial symmetry group in its transformation semigroup. Statistical complexity: 4.518, entropy rate: 0.570709.  The $\epsilon$-machine for the non-meditator has 43 causal states, transition semigroup with holonomy decomposition length 23 and Krohn-Rhodes complexity $0$, indicating 
no non-trivial symmetry groups in its transformation semigroup. Statistical complexity: 3.5429;  entropy rate: 0.50769.
}
 \label{Meditators}
 \end{figure}
 


Our prediction was that meditators will have  significantly higher number of causal states in $\epsilon$-machines than non-meditators in mind-wandering and open presence modes due to less fixation on the current mental state and greater experiential openness/less dependence on automatic habitual patterns of reaction/response to stimulus/mental content. EEG microstate observations and hence transitions were sampled at 250 Hz (i.e.\ each transition is 4 ms) for each mind-mode  over two-minute intervals. EEG microstate classes used were subject-specific rather than group-averaged as in EEG microstates classes derived by Koenig et al. \cite{Koenig2002}, so the transition probabilities and $\epsilon$-machines reported here were derived at subject-level during a single 2-minute interval rather than group-level. 
Figure~\ref{Meditators} displays reconstructed $\epsilon$-machines from EEG microstate sequences from one experienced meditator and
one non-meditator during mind-wandering (or `neutral' resting state).

As can be seen from the figure, EEG microstate sequences considering only the current microstate would be a very poor fit to the actual data, as dozens or hundreds of causal states may be required to describe the dynamics and syntax in either meditators or meditation-na\"{\i}ve controls. The results of our pilot study  generally showed the number of causal states to be significantly higher for meditators than meditation-na\"{\i}ve controls \cite{GwynThesis} and appear to support the hypothesis that meditators' neurodynamics as indexed by EEG microstates are more complex than those of controls, but this needs to be confirmed/replicated with more detailed studies currently underway.

\newpage
\section{Summary and Conclusions}\label{Conc}
EEG microstates, although unlikely to literally translate into ``the atoms of thought" as proposed by Lehmann et al \cite{Lehmann1998} due to the limited variability of their spatial topographies, have been shown to correspond to changes in the type of mental content (e.g. visual imagery vs.\ abstract thought), information processing dynamics (e.g. speed of information processing, degree of conceptual elaboration/mode of attention), as well as levels of wakefulness/alertness. They are therefore most likely to reflect global neural integration processes. Spatiotemporal dynamics or sequences of four microstate classes have been shown to differentiate between clinical populations and healthy controls. Thus far, observational and model-fitting studies of microstate syntax have only considered the current microstate to determine the probability of the next one.  However, the neural dynamics and thus EEG microstate syntax and grammar are likely to be far more complex. We therefore propose the application of $\epsilon$-machines to EEG microstate sequences as a method capable of reconstructing much greater complexity of the underlying neurodynamics processes. 
 
 The experiments reported here using simulated data validate the capacity of $\epsilon$-machine to reconstruct the pre-given structure. Furthermore, applying $\epsilon$-machines to data reported by Lehmman et al. \cite{Lehmann2005} showed that a simple transition structure could account for predominance of certain reported sequences in schizophrenia patients vs. healthy controls, but also suggested that more subtle dynamical structure could be uncovered using the actual sequences. 
 Different ways of modeling the data (individual vs. group/population) and their relative value can be addressed using the $\epsilon$-reconstruction method which allows the use
 of multiple sequences:  Subject-level EEG microstate analysis for individuals in particular mental states, with particular expertise or medical conditions, or doing particular tasks in specific contexts can be tailored  to detect and address the personal characteristics and needs of particular individuals.   Comparing such {\em subject-specific} models to $\epsilon$-machines reconstructed  at {\em group-level} by combining data from all members of a particular population or clinical group could help establish and assess the value of subject-level analysis over group-level EEG microstates classes/analysis in various domains of application.
 
Applied to EEG microstate sequences from expert meditators vs. healthy controls during 2-minute periods of different mind modes, $\epsilon$-machines revealed syntactic structure of the data beyond what can be produced by simple transition systems that depend only on the current microstate. Single case studies presented here (Section~\ref{Med}) and significant group analysis results of the pilot study \cite{GwynThesis} suggest that the number of causal states in the $\epsilon$-machines may generally be higher in expert meditators than controls, but these preliminary results need to be replicated in further studies, which are currently underway.
  These observations tentatively suggest examining syntactic structure of microstate sequences will be required to understand the underlying neurodynamics. 
Furthermore, our results using $\epsilon$-machines to derive the sequence structure of the resting EEG microstate data from experienced meditators and meditation-na\"{\i}ve controls support the need for modeling EEG microstates with far greater complexity than applied in previous research.  Overall, our results suggest that task, modality, and inter-individual differences in neural dynamics may be reflected in EEG microstate syntax that one could capture using $\epsilon$-machine analysis at a subject level. 
Further research is needed to fully tap into the potential of the $\epsilon$-machine method for the study of spatiotemporal neural dynamics and its application in health and disease.

\section*{Acknowledgment}
\noindent
This work is supported in part by a BIAL Foundation grant to the authors
for the project ``Decoding the Language of `Now': EEG microstates in experienced meditators, from letters to grammar''.

\newpage



\begin{thebibliography}{10}
\providecommand{\url}[1]{#1}
\csname url@samestyle\endcsname
\providecommand{\newblock}{\relax}
\providecommand{\bibinfo}[2]{#2}
\providecommand{\BIBentrySTDinterwordspacing}{\spaceskip=0pt\relax}
\providecommand{\BIBentryALTinterwordstretchfactor}{4}
\providecommand{\BIBentryALTinterwordspacing}{\spaceskip=\fontdimen2\font plus
\BIBentryALTinterwordstretchfactor\fontdimen3\font minus
  \fontdimen4\font\relax}
\providecommand{\BIBforeignlanguage}[2]{{%
\expandafter\ifx\csname l@#1\endcsname\relax
\typeout{** WARNING: IEEEtran.bst: No hyphenation pattern has been}%
\typeout{** loaded for the language `#1'. Using the pattern for}%
\typeout{** the default language instead.}%
\else
\language=\csname l@#1\endcsname
\fi
#2}}
\providecommand{\BIBdecl}{\relax}
\BIBdecl

\bibitem{Hokusai}
K.~Hokusai, ``Under a {W}ave off {K}anegawa,'' in \emph{Thirty Six Views of
  Mount Fuji}, 1829-1833, [Public Domain Image].

\bibitem{Lehmann2005}
D.~Lehmann, P.~L. Faber, S.~Galderisi, W.~M. Herrmann, T.~Kinoshita,
  M.~Koukkou, A.~Mucci, R.~D. {Pascual-{M}arqui}, N.~Saito, J.~Ackermann,
  G.~Winterer, and T.~Koenig, ``{E}{E}{G} microstate duration and syntax in
  acute, medication-naïve, first-episode schizophrenia: a multi-center
  study,'' \emph{Psychiatry Research: Neuroimaging}, vol. 138, no.~2, p.
  141–156, 2005.

\bibitem{Nishida2013}
K.~Nishida, Y.~Morishima, M.~Yoshimura, T.~Isotani, S.~Irisawa, K.~Jann,
  T.~Dierks, W.~Strik, T.~Kinoshita, and T.~Koenig, ``{E}{E}{G} microstates
  associated with salience and frontoparietal networks in frontotemporal
  dementia, schizophrenia and {A}lzheimer’s disease,'' \emph{Clincial
  Neurophysiology}, vol. 124, no.~6, p. 1106–1114, 2013.

\bibitem{CrutchfieldYoung1989}
J.~P. Crutchfield and K.~Young, ``Inferring statistical complexity,''
  \emph{Physical Review Letters}, vol.~62, no.~2, pp. 105--108, 1989.

\bibitem{Crutchfield1993}
J.~P. Crutchfield, ``Observing complexity and the complexity of observation,''
  in \emph{Inside versus Outside}, H.~Atmanspacher, Ed.\hskip 1em plus 0.5em
  minus 0.4em\relax Springer Verlag, 1993, pp. 235--272.

\bibitem{James}
W.~James, \emph{Principles of Psychology}.\hskip 1em plus 0.5em minus
  0.4em\relax Cambridge, 1981 [1890].

\bibitem{Lehmann1998}
D.~Lehmann, W.~K. Strik, B.~Henggeler, T.~Koenig, and M.~Koukkou, ``Brain
  electrical microstates and momentary conscious mind states as building blocks
  of spontaneous thinking: I. visual imagery and abstract thoughts,'' \emph{Int
  J Psychophysiol}, vol.~29, pp. 1--11, 1998.

\bibitem{Lehmann1987}
D.~Lehmann, H.~Ozaki, and I.~Pal, ``{E}{E}{G} alpha map series: Brain
  micro-states by space-oriented adaptive segmentation,'' \emph{Electroencheph
  Clin Neurophysiol}, vol.~67, pp. 271--288, 1987.

\bibitem{Koenig2002}
T.~Koenig, L.~Prichep, D.~Lehmann, P.~{Vadles Sosa}, E.~Braeker, H.~Kleinlogel,
  R.~Isenhart, and E.~R. John, ``Milisecond by milisecond, year by year:
  Normative {E}{E}{G} microstates and developmental stages,''
  \emph{NeuroImage}, vol.~16, pp. 41--48, 2002.

\bibitem{Kindler2011}
J.~Kindler, D.~Hubl, W.~Strik, T.~Dierks, and T.~Koenig, ``Resting-state
  {E}{E}{G} in schizophrenia: auditory verbal hallucinations are related to
  shortning of specific microstates,'' \emph{Clin Neurophysiol}, vol. 122, pp.
  1179--1182, 2011.

\bibitem{Koenig1999}
T.~Koenig, D.~Lehmann, M.~C. Merlo, K.~Kochi, D.~Hell, and M.~Koukkou, ``A
  deviant {E}{E}{G} brain microstate in acute, neuroleptic-na\"{\i}ve
  schizophrenics at rest,'' \emph{Eur Arch Psychiatry Clin Neurosci}, vol. 249,
  no.~4, pp. 205--211, 1999.

\bibitem{Strik1995}
W.~Strik, T.~Dierks, T.~Becker, and D.~Lehmann, ``Larger topogrpahical variance
  and decreased duration of brain electric microstates in depression,'' \emph{J
  Neural Transm Gen Sect JNT}, vol.~99, pp. 213--222, 1995.

\bibitem{Dierks1997}
T.~Dierks, V.~Julin, K.~Maurer, L.~O. Wahlund, O.~Almkvist, W.~K. Strik, and
  B.~Winblad, ``{E}{E}{G}-microstates in mild memory impairment and alzheimer's
  disease: Possible assocation with distrubed information processing,'' \emph{J
  Neural Transm}, vol. 104, pp. 483--495, 1997.

\bibitem{Brodbeck2012}
V.~Brodbeck, A.~Kuhn, F.~{von Wegner}, A.~Morzelewski, E.~Tagliazucchi,
  S.~Borisov, C.~M. Michel, and H.~Laufs, ``{E}{E}{G} microstates of
  wakefulness and {N}{R}{E}{M} sleep,'' \emph{NeuroImage}, vol.~62, no.~3, pp.
  2129--39, 2012.

\bibitem{Schmidtke2004}
J.~I. Schmidtke and W.~Heller, ``Personality, affect and {E}{E}{G}: predicting
  patterns of regional brain activity related to extraversion and
  neuroticism,'' \emph{Personal Individ Differ}, vol.~36, pp. 717--732, 2004.

\bibitem{Katayama2007}
H.~Katayama, R.~Gianotti, T.~Isotani, P.~L. Faber, K.~Sasada, T.~Kinoshita, and
  D.~Lehmann, ``Classes of multichannel {E}{E}{G} microstates in light and deep
  hypnotic conditions,'' \emph{Brain Topogr}, vol.~20, pp. 7--14, 2007.

\bibitem{Britz2010}
J.~Britz, D.~{Van De Ville}, and C.~M. Michel, ``{B}{O}{L}{D} correlates of
  {E}{E}{G} topography reveal rapid resting-state network dynamics,''
  \emph{NeuroImage}, vol.~52, pp. 1162--1170, 2010.

\bibitem{Milz2016}
P.~Milz, P.~L. Faber, D.~Lehmann, T.~Koenig, K.~Kochi, and R.~D.
  {Pascual-{M}arqui}, ``The functional significance of {E}{E}{G}
  microstates-associations with modalities of thinking,'' \emph{NeuroImage},
  vol. 125, pp. 643--656, 2016.

\bibitem{Wackermann1993}
J.~Wackermann, D.~Lehmann, C.~M. Michel, and W.~K. Strik, ``Adaptive
  segmentation of spontaneous {E}{E}{G} map series into spatially defined
  microstates,'' \emph{Int J Psychophysiol}, vol.~14, p. 269–283, 1993.

\bibitem{Schlegel2012}
F.~Schlegel, D.~Lehmann, P.~L. Faber, P.~Milz, and L.~R. Gianotti, ``{E}{E}{G}
  microstates during resting represent personality differences,'' \emph{Brain
  Topogr}, vol.~25, pp. 20--26, 2012.

\bibitem{VanDeVille2010}
D.~{Van {D}e {V}ille}, J.~Britz, and C.~M. Michel, ``{E}{E}{G} microstate
  sequences in rest reveal scale-free dynamics,'' \emph{Proc Natl Acad Sci},
  vol. 107, no.~42, p. 18179–18184, 2010.

\bibitem{Hopcroft}
J.~E. Hopcroft, R.~Motwani, and J.~D. Ullman, \emph{Introduction to Automata,
  Languages, and Computation}, 3rd~ed.\hskip 1em plus 0.5em minus 0.4em\relax
  Pearson, 2006.

\bibitem{EilenbergB}
S.~Eilenberg, \emph{Automata, Languages and Machines}.\hskip 1em plus 0.5em
  minus 0.4em\relax Academic Press, 1976, vol.~B.

\bibitem{Nehaniv1996}
C.~L. Nehaniv, ``Complexity of finite aperiodic semigroups and star-free
  languages,'' in \emph{Semigroups, Automata, Languages}, J.~Almeida, G.~Gomes,
  and P.~Silva, Eds.\hskip 1em plus 0.5em minus 0.4em\relax World Scientific,
  1996, pp. 195--209.

\bibitem{wildbook}
J.~Rhodes, \emph{{Applications of Automata Theory and Algebra via the
  Mathematical Theory of Complexity to Biology, Physics, Psychology,
  Philosophy, and Games}}.\hskip 1em plus 0.5em minus 0.4em\relax World
  Scientific Press, 2010, foreword by M.~W.~Hirsch, ed.: C.~L.~Nehaniv.

\bibitem{Gaertner2010}
M.~G\"{a}rtner, V.~Brodbeck, H.~Laufs, and G.~Schneider, ``A stochastic model
  for {E}{E}{G} microstate sequence analysis,'' \emph{NeuroImage}, vol. 104,
  pp. 199--208, 2010.

\bibitem{Shalizi2003}
C.~R. Shalizi, K.~L. Shalizi, and J.~P. Crutchfield, ``An algorithm for pattern
  discovery in time series,'' \emph{J. Machine Learning Research, arXiv
  preprint cs/0210025}, 2003.

\bibitem{Shalizi2004}
C.~R. Shalizi and K.~L. Klinkner, ``Blind construction of optimal nonlinear
  recursive predictors for discrete sequences,'' in \emph{Uncertainty in AI:
  Proc. 20th Conf.}\hskip 1em plus 0.5em minus 0.4em\relax AUAI Press, 2004,
  pp. 504--511.

\bibitem{royalsoc}
C.~L. Nehaniv, J.~Rhodes, A.~{Egri-Nagy}, P.~Dini, E.~{Rothstein Morris},
  G.~Horv{\'a}th, F.~Karimi, D.~Schreckling, and M.~J. Schilstra, ``Symmetry
  structure in discrete models of biochemical systems: natural subsystems and
  the weak control hierarchy in a new model of computation driven by
  interactions,'' \emph{Phil. Trans. R. Soc. A}, vol. 373, no. 2046, p.
  20140223, 2015.

\bibitem{AttilaJamesChrystopher}
A.~Egri-Nagy, J.~D. Mitchell, and C.~L. Nehaniv, ``Sgp{D}ec: Cascade
  (de)compositions of finite transformation semigroups and permutation
  groups,'' in \emph{International Congress on Mathematical Software}, vol.
  8592 Lecture Notes in Computer Science.\hskip 1em plus 0.5em minus
  0.4em\relax Springer Verlag, 2014, pp. 75--82.

\bibitem{GwynThesis}
G.~Jones, ``{E}{E}{G} microstates in meditators,''  MSc Thesis
  in Neuroscience - Supervised by E. Antonova, J. Nottage and C. L. Nehaniv,
  King's College London, 2013.

\end{thebibliography}

\end{document}